\documentclass[12pt]{article}
\usepackage{amsfonts}
\usepackage{amssymb}

\def \id {1\!\!{\rm I}}

\begin{document}
\hspace{12.5cm}{\textbf{ETH-TH/97-39}}

\hspace{12.5cm}{

\vspace*{1cm}
\centerline{\large{\bf{Absence of Zero Energy States in Reduced}}}
\centerline{\large{\bf{SU(N) 3d Supersymmetric Yang Mills Theory}}}
\vspace{1.5cm}
\centerline{Jens Hoppe}
\centerline{Theoretische Physik, ETH H\"onggerberg}
\centerline{CH-8093 Z\"urich}
\vspace{0.5cm}
\centerline{Shing-Tung Yau}
\centerline{Mathematics Department, Harvard University}
\centerline{Cambridge, MA 02139}
\vspace{1.5cm}
\centerline{\large Abstract}
\vspace{0.5cm}
\noindent For the SU(N) invariant supersymmetric matrix model related to
membranes in 4 space-time dimensions we argue that 
$\langle\Psi,\chi\rangle = 0$ for the previously
obtained solution of $Q\chi=0, Q^{\dagger}\Psi=0$.

\vfill\eject  
In a series of 3 short papers [1] it was recently shown how to obtain, for a
certain class of supersymmetric matrix models, solutions of $Q\psi=0$
resp. $Q^{\dagger}\psi=0$. The models of interest [2] are SU(N)
gauge-invariant and can be formulated with either 2, 3, 5 or 9 times
$(N^2-1)\cdot 2$ bosonic degrees of freedom. This letter mainly concerns the
first case, $d=2$ (corresponding to membranes in 4 space-time dimensions
[3], and to 2+1 dimensional (Susy) Yang-Mills theory with spatially constant
fields [3]), while our method
can also be applied to the other cases. For different approaches
to the problem see [4],[5],[6],[7].

The supercharges of the model are given by
\begin{eqnarray}
Q   & = & iq_a\lambda_a + 2\partial_a\frac{\partial}{\partial\lambda_a}=:
M_a\lambda_a+D_a\partial_{\lambda{_a}}\nonumber\\  
Q^{\dagger} & = & - i q_a\frac{\partial}{\partial\lambda_a} - \,
2\overline{\partial}_a\lambda_a=:M_a^{\dagger}\partial_{\lambda{_a}}+D_a^{\dagger}\lambda_a
\end{eqnarray}
where $\partial_a=\frac{\partial}{\partial z_a}, z_a \ \epsilon \ {\mathbb C},
a=1\cdots N^2-1, q_a:=\frac{i}{2} f_{abc}z_b\overline{z}_c$ ($f_{abc}$ being
totally antisymmetric, real, structure constants of SU(N)) and $\lambda_a$
$(\frac{\partial}{\partial\lambda_a})$ being fermionic creation (annihilation)
operators satisfying $\{\lambda_a, \
\frac{\partial}{\partial\lambda_b}\}=\delta_{ab} \ , \ \{\lambda_a ,
\lambda_b\}=0=\{\frac{\partial}{\partial\lambda_a} \ , \
\frac{\partial}{\partial\lambda_b}\}$. In ${\cal{H}}_+$, the Hilbert-space of
SU(N)-invariant square-integrable wavefunctions
\begin{equation}
\Psi \ = \ \psi \ + \ \frac{1}{2} \psi_{ab}\lambda_a\lambda_b \ + \ \cdots \ +
\ \frac{1}{\Lambda !}\psi_{a{_1}\cdots
  a{_{\Lambda}}}\lambda_{a{_{1}}}\cdots\lambda_{a{_{\Lambda}}} \ \ ,
\end{equation}
$\Lambda := \ N^2-1$ (even) the general solution of 
\begin{equation}
Q^{\dagger}\Psi \ = \ 0 \ , \ Q \chi \ = \ 0
\end{equation}
was shown [1] to be of the form
\begin{eqnarray}
\Psi & = & ({\id}-A)^{-1} \ \Psi^{(h)} \nonumber \\
\chi & = & ({\id}-B)^{-1} \ \chi^{[h]}
\end{eqnarray}
with
\begin{eqnarray}
A:= \ (I^{\dagger}\cdot \lambda)( D^{\dagger}\cdot \lambda )\ , \
B & = & (I\cdot\partial_{\lambda})(D\cdot \partial_{\lambda}) \nonumber \\
I_a: & = & i\frac{q_a}{q^2}
\end{eqnarray}
and
\begin{equation}
(M^{\dagger}\cdot\partial_{\lambda}) \ \Psi^{(h)} = 0 \ , \ (M\cdot\lambda) \
\chi^{[h]}= 0 \ .
\end{equation}
As
\begin{equation}
\Psi = \Psi^{(h)} + A\Psi \ , \ \chi = \chi^{[h]} + B\chi \ ,
\end{equation}
and (from (3))
\begin{eqnarray}
A\Psi & = & - \ (I^{\dagger}\cdot\lambda) \ (M^{\dagger}\partial_{\lambda}) \
\Psi = \frac{q_aq_b}{q^2}\lambda_a\partial_{\lambda{_b}}\Psi \ \ \epsilon \ \
{\cal{H}}_+ \\ 
B\chi & = & - \ (I \cdot \partial_{\lambda}) \ (M\cdot\lambda) \ \chi =
\chi-\frac{q_aq_b}{q^2} \lambda_a\partial_{\lambda{_b}}\chi \ \
\epsilon \ \ {\cal{H}}_+ \nonumber
\end{eqnarray}
one can see that $\Psi^{(h)}$, resp. $\chi^{[h]}$, have to be elements of
${\cal{H}}_+$. The scalar product of any two solutions of (3) is therefore
\begin{eqnarray}
\langle \Psi , \chi \rangle & = & \langle (\id -A)^{-1}\Psi^{(h)} \ , \
(\id-B)^{-1} \chi^{[h]}\rangle \nonumber \\
 & = & \langle (\id -B^{\dagger})^{-1}(\id-A)^{-1} \  \Psi^{(h)} \ , \
 \chi^{[h]} \rangle 
 \nonumber \\
 & = & \langle (\id-C)^{-1} \Psi^{(h)} \ , \ \chi^{[h]}\rangle
\end{eqnarray}
with 
\begin{equation}
C := A+B^{\dagger} = \{ I^{\dagger} \cdot \lambda , D^{\dagger} \cdot \lambda
\} \ .
\end{equation}
As $H_M:= \{ M \cdot \lambda , M^{\dagger} \cdot \partial_{\lambda} \} = q^2 >
0$, (6) implies
\begin{equation}
\Psi^{(h)} = (M^{\dagger}\cdot\partial_{\lambda}) \Psi_{-}^{(h)},
\chi^{[h]}=(M\cdot\lambda)\chi_{-}^{[h]}
\end{equation}
for some $\Psi^{(h)}_{-}, \chi^{[h]}_{-}$.

Furthermore, $C$ commutes with $M^{\dagger}\cdot\partial_{\lambda}$ (between
SU(N) invariant states), as 
\begin{equation}
[M^{\dagger}\cdot\partial_{\lambda} \ , \ \{ I^{\dagger}\cdot\lambda ,
D^{\dagger}\cdot\lambda \}] = -[I^{\dagger}\cdot{\lambda} ,
  \{  D^{\dagger}\cdot\lambda , M^{\dagger}\cdot\partial_{\lambda} \}] -
  [D^{\dagger}\cdot\lambda , \{ M^{\dagger}\cdot\partial_{\lambda},
  I^{\dagger}\cdot \lambda \}]
\end{equation}
and $\{D^{\dagger}\cdot\lambda , M^{\dagger}\cdot\partial_{\lambda} \}=
-iz_aJ_a$ ,
\begin{equation}
J_a:= -i f_{abc}(z_b\partial_c+\bar{z}_b\bar{\partial}_c+\lambda_b\partial_{\lambda{_c}})
\ .
\end{equation}
One therefore has 
\begin{equation}
\langle \Psi , \chi \rangle = \langle
(M^{\dagger}\cdot\partial_{\lambda}) \ (1-C)^{-1}\Psi^{(h)}_{-} \ , \
(M\cdot\lambda) \ \chi^{[h]}_{-}\rangle = 0,
\end{equation}

showing that
$Q^{\dagger}\Psi=0=Q\Psi$ implies $\Psi\equiv 0 $ (in ${\cal{H}}_+$). The same
holds in ${\cal{H}}_{-}$ (as the extra-conditions
$D^{\dagger}\cdot\lambda\psi^{(h)}_{a{_1}}\cdots_{a{_{\Lambda-1}}}\lambda_{a{_1}}\cdots\lambda_{a{_{\Lambda-1}}}=0
\ , \ (D\cdot\partial_{\lambda}) \ \chi^{[h]}_a\lambda_a = 0 $ , are
automatically satisfied for SU(N)-invariant states, due to (11)). 

Let us close
with a remark on d=9: in order to prove the existence of a zero-energy
state for the supersymmetric matrix model related to membranes in 11
space-time dimensions it is sufficient to show that for one particular
solution of $Q^{\dagger}\Psi=0$, and one particular solution of $Q\chi =0$,
one has $\langle\Psi ,\chi\rangle\neq0$. 

{\it Note added:} 
Due to the singularity at $q = 0$ the above argument is not yet
complete.

\vspace{0.2cm}
\centerline{\large References}
\vspace{0.2cm}

\begin{description}
\item[[1]] J.~Hoppe; hep-th/9709132/9709217/9711033.
\vspace{-0.3cm}
\item[[2]] J.~Goldstone, J.~Hoppe; unpublished (resp. [3]).
\vspace{-0.3cm}
\item[] M.~Baake, P.~Reinicke, V.~Rittenberg; Journal of Math. Physics 26
  (1985) 1070.
\vspace{-0.3cm}
\item[] M.~Claudson, M.~Halpern; Nuclear Physics B 250 (1985) 689.
\vspace{-0.3cm}
\item[] R.~Flume; Annals of Physics 164 (1985) 189. 
\vspace{-0.3cm}
\item[] J.~Hoppe; in ``Constraint's Theory and Relativistic Dynamics'', World
  Scientific 1987.

\vspace{-0.3cm}
\item[] B.~de Wit, J.~Hoppe, H.~Nicolai; Nuclear Physics B305 (1988) 545.
\vspace{-0.3cm}
\item[] T.~Banks, W.~Fischler, S.H.~Shenker, L.~Susskind; hep-th/9610043.
\vspace{-0.3cm}
\item[[3]] J.~Hoppe; ``Quantum Theory of a Massless Relativistic Surface'', MIT
  Ph.D Thesis 1982.
\vspace{-0.3cm}
\item[[4]] J.~Fr\"ohlich, J.~Hoppe; hep-th/9701119.
\vspace{-0.3cm}
\item[[5]] P.~Yi; hep-th/9704098.
\vspace{-0.3cm}
\item[[6]] S.~Sethi, M.~Stern; hep-th/9705046.
\vspace{-0.3cm}
\item[[7]] M.~Porrati, A.~Rozenberg; hep-th/9708119.
\vspace{-0.3cm}

\end{description}
\end{document}